\documentstyle[12pt]{article}
\input epsf.sty
\newcommand{\be}{\begin{equation}}
\newcommand{\ee}{\end{equation}}
\newcommand{\bea}{\begin{eqnarray}}
\newcommand{\eea}{\end{eqnarray}}
\newcommand{\al}{\alpha}
\newcommand{\bt}{\beta}
\newcommand{\gm}{\gamma}
\newcommand{\Gm}{\Gamma}
\newcommand{\dl}{\delta}

\newcommand{\eps}{\epsilon}

\newcommand{\th}{\theta}

\newcommand{\kp}{\kappa}

\newcommand{\nn}{\nonumber}

\newcommand{\cA}{{\cal A}}
\newcommand{\cB}{{\cal B}}
\begin{document}
\title{String Thermalization in Static Spacetimes}
\author{A. Kuiroukidis\thanks{E-mail address: kuirouki@astro.auth.gr}, 
D. B. Papadopoulos\thanks{E-mail address: papadop@astro.auth.gr} and 
J. E. Paschalis\thanks{E-mail address: paschalis@physics.auth.gr} \\
Department of Physics, \\
\small Section of Astrophysics, Astronomy and Mechanics, \\
\small Aristotle University of Thessaloniki, \\
\small 54006 Thessaloniki, GREECE}

\maketitle
\begin{abstract}
We study the evolution, the transverse spreading 
and the subsequent thermalization of string states 
in the Weyl static axisymmetric spacetime. 
This possesses a singular event horizon on the symmetry 
axis and a naked singularity along the other directions. 
The branching diffusion process of string 
bits approaching the singular black-hole horizon provides 
the notion of the temperature that is calculated for this process. 
We find that the solution of the Fokker-Planck equation in the 
phase space of the transverse variables of the string, can be 
factored as a product of two thermal distributions, provided 
that the classical conjugate variables satisfy the 
uncertainty principle. 
We comment on the possible physical significance of this result. 

\end{abstract}

\section*{I. Introduction}\
String theory is the most promising candidate for a consistent 
quantization of gravity and a subsequent unified 
description of all the fundamental interactions [1-2]. One 
of the first steps towards a deep understanding of the quantum 
gravitational phenomena is to study the string  evolution and
quantization in the presence of a gravitational field [2-3]. This 
became necessary by the fact that in the spectrum of bosonic 
string theory gravity appears naturally through the massless 
spin-2 state having long range couplings that mimic General 
Relativity. 

In the context of Black Hole Physics, strings have been identified 
with black hole states [4] and also have been used for the 
resolution of the so called {\sl Information Loss} problem. Namely 
of the apparent inconsistency of the black hole evaporation through 
the thermal Hawking radiation and the unitary evolution of 
the quantum states of the infalling matter that produces the 
gravitational collapse. The notion of Stretched Horizon 
has been introduced for the distant observer, and it is supposed to 
absorb and 
thermalize the quantum states of infalling matter which is in the form 
of strings [5-7]. 

Following these arguments, the transverse spreading of a relativistic 
string which falls towards the Black Hole Horizon has been described 
as a {\sl branching diffusion} process [8]. This stohastic process 
provides the necessary mechanism for thermalization of the quantum 
state of the string. The resulting temperature calculated agrees 
in the order of magnitude, with the semiclassical result of 
Hawking and Bekenstein. 
Also, other efficient methods have been proposed that use explicit 
equations of state for the matter in the form of strings, the so called 
Planckian solid, which prevent the loss of information inside the 
black hole during the gravitational collapse [9].

All the above efforts have been concentrated on the problem where
the singularity is hidden from the asymptotic observer via the Horizon. 
It would be interesting to discuss the problem of whether 
thermalization of string states can occur in a more general context 
where there exists a singular event horizon or a 
naked singularity in the spacetime manifold. 
In principle one can get arbitrary close to the singularity in this case. 
However, if one takes into account the fact that strings are 
quantum objects, then one can show that 
there exists thermalization of string states, 
occuring at finite distance from the singularity. 
The purpose of this paper is to examine this possibility. 

The Weyl Static Axisymmetric Spacetime is the model under consideration 
that provides the setting and it is a class of Static exact solutions 
of the Einstein's field 
equations [10]. The problem of a line source of length $\; 2\al \; $
and of mass density $\; \gm /2\; $is known to be described by the 
$\; \gm \; $metric, which can be written either in spherical or 
prolate spheroidal coordinate systems.
This family of solutions encompasses the Schwarzschild solution for
$\; \gm =1\; $ and except for that solution, the family possesses
singular event horizons [11]. These are in fact directional singularities
for $\;\gm \geq 2\; $, 
and by a proper choice of the coordinate system reveal their nature as
extended hypersurfaces [12]. For all the other values of $\gm $ one 
in fact has a singularity for all directions$\; 0\leq \th \leq \pi \; $[12]. 
From this the $\; \gm _{A}\; $metric
which is an exact solitonic solution in vacuum 
is obtained through a limiting procedure [13].
This can also be interpreted as the metric for a counter-rotating 
disc in General Relativity [14,15]. 

This paper is organized as follows: \\ 
In section II, the general features of Static Axisymmetric 
solutions to the Einstein's field equations are reviewed. \\
In section III, we develop the formalism and calculate the 
diffusion coefficients for the thermalization process 
of the string states, in the case where 
$\; 0<\gm <1\; $, which is the range of parameters under 
consideration. \\
In section IV, The Fokker-Planck equation is solved in the phase 
space of the transverse variables of the string, that is falling 
towards the black-hole. \\
In section V, numerical estimates that connect the 
above work with real astrophysical systems, is provided.\\
In section VI, we give a discussion of the contribution of 
the quantum aspects of strings, when they are taken into account, 
for the process of thermalization. \\ 
\section*{II. Static Axisymmetric Black-Hole Spacetimes}\
The curved spacetime manifold that we want to study
is given by the Weyl metric [11], 
\bea
ds^{2}&=&-\left(\frac{\xi -1}{\xi +1}\right)^{\gamma }dt^{2}+
\alpha ^{2}\left(\frac{\xi ^{2}-1}{\xi ^{2}-\eta ^{2}}\right)^{\gamma ^{2}}
\left(\frac{\xi +1}{\xi -1}\right)^{\gamma }(\xi ^{2}-\eta ^{2})\cdot \nn \\
&\cdot &\left[\frac{d\xi ^{2}}{\xi ^{2}-1}+\frac{d\eta ^{2}}{1-\eta ^{2}}\right]+
\alpha ^{2}\left(\frac{\xi +1}{\xi -1}\right)^{\gamma }
(\xi ^{2}-1)(1-\eta ^{2})d\phi ^{2}, 
\eea
written in prolate, spheroidal coordinates, 
$\; (t,\xi ,\eta ,\phi )\; \; $. 
The transformation
from the Cartesian coordinates, 
is given by 
\bea
x^{2}&=&\alpha ^{2}(\xi ^{2}-1)(1-\eta ^{2})cos^{2}\phi \\
y^{2}&=&\alpha ^{2}(\xi ^{2}-1)(1-\eta ^{2})sin^{2}\phi \\
z&=&\alpha \xi \eta ,
\eea
where $\; \; \alpha ,\gamma \; \; $ are constants, 
representing 
a line source of length $\; 2\alpha \; $ and a mass density
$\; \gamma /2\;$.
The range of the prolate spheroidal coordinates is given by \\
$\; -\infty <t<+\infty ,\; \; 
1\leq \xi <+\infty ,\; \; 
-1\leq \eta \leq +1,\; \; 
0\leq \phi <2\pi .\; $

The $\; \gamma \; $metric in the $\; (t,r,\th ,\phi )\; $ 
coordinate system, using  
the transformation 
\bea
\rho ^{2}&=&(r^{2}-2mr)sin^{2}\theta \\
z&=&(r-m)cos\theta ,
\eea
reads 
\bea
ds^{2}=-{\cal A}^{\gm }dt^{2}+\frac{{\cal A}^{\gm ^{2}-\gm -1}}
{{\cal B}^{\gm ^{2}-1}}
dr^{2}+
\frac{{\cal A}^{\gm ^{2}-\gm }}{{\cal B}^{\gm ^{2}-1}}r^{2}d\th ^{2}
+{\cal A}^{1-\gm }r^{2}sin^{2}\th d\phi ^{2},
\eea
where 
\bea
{\cal A}&=&(1-\frac{2m}{r})\\
{\cal B}&=&(1-\frac{2m}{r}+\frac{m^{2}}{r^{2}}sin^{2}\th ).
\eea 
The null outgoing vector is given by
\bea
l^{\al }&=&\left(-{\cal A}^{-\gm },
\frac{{\cal B}^{(\gm ^{2}-1)/2}}{{\cal A}^{(\gm ^{2}-1)/2}},0,0\right)\\
l_{\al }&=&\left(1,
\frac{{\cal A}^{(\gm ^{2}-2\gm -1)/2}}{{\cal B}^{(\gm ^{2}-1)/2}},0,0\right)\\
l^{\al }l_{\al }&=&0.
\eea
Introducing the timelike Killing vector 
$\; \; \xi ^{\mu }_{(t)}=\dl ^{\mu }_{t}\; \; $, 
we have 
$\; \; \xi ^{\al }l_{\al }=1\; \; $. 

The surface gravity is defined as 
$\; \; \kp =l^{\al }\xi ^{b}\nabla _{b}\xi _{\al }\; \; $[11]. The result is 
\be
\kp =m\frac{1}{r^{2}}{\cal A}^{(\gm ^{2}-1)/2}{\cal B}^{-(\gm -1)^{2}/2}.
\ee 
We have the following two cases, when $\; r \rightarrow 2m\; $.

For $\; \th \neq 0, \pi \; $ when $\; \; \gm \neq 1\; \; $
the surface gravity tends to infinity. 
For $\; \th =0, \pi \; $ we have
\be
\kp =\frac{m}{r^{2}}(1-\frac{2m}{r})^{\gm -1} 
\ee 
and it tends to zero when $\; \gm >1\; $, while it tends to infinity 
as 
$\; 0<\gm <1\; $. \\
Therefore, intuitively one expects that in 
the former class of models it is difficult to have thermalization 
process while in the second it can occur naturally. This is because 
in the first case the strong tidal gravitational forces tend to shrink 
the string, so it is impossible for the string bits to become 
uncorrelated (via thermalization process) as the 
relativistic string approaches the singular event horizon. \\
\section*{III. String Thermalization on the Symmetry Axis.}\ 
The $\; \gm -$metric for $\; \th =0,\; \pi \; $is given by
\be
ds^{2}=-\left(1-\frac{2m}{r}\right)^{\gm }dt^{2}
+\left(1-\frac{2m}{r}\right)^{-\gm }dr^{2}=
-\left(1-\frac{2m}{r}\right)^{\gm }dudv, 
\ee
where 
\bea
du&=&dt+\frac{dr}{\left(1-\frac{2m}{r}\right)^{\gm }}\\
dv&=&dt-\frac{dr}{\left(1-\frac{2m}{r}\right)^{\gm }}. 
\eea
We introduce the new variables $\; (s,p)\; $,
by 
\bea
du&=&4m(ds/cos^{\gm }(s)) \\
dv&=&4m(dp/sin^{\gm }(p)). 
\eea 
The coefficients have been chosen in such a way so as to reproduce 
the well known result [8] 
in the case of Schwarzschild Spacetime, 
as will be shown below. 

The metric assumes the Rindler-type form  
\bea
ds^{2}&=&-4Q^{2}dsdp\\
Q^{2}&=&\frac{4m^{2}\left(1-\frac{2m}{r}\right)^{\gm }}
{cos^{\gm }(s)sin^{\gm }(p)}. 
\eea
where $\; Q|_{r=2m}\equiv \sqrt{A}\; $ is finite, 
continuous and non-zero on the horizon [12]. \\ 
Integration of Eqs (18), (19) gives $\; (0<\gm <1)\; $, \\
(Appendix II) 
\bea 
u(s)&=&\frac{4m}{1-\gm }cos^{1-\gm }(s)\; \; _{2}F_{1}
\left(1-\gm ,1;\frac{3-\gm }{2};\frac{1}{2}[1+sin(s)]\right)\\
v(p)&=&\frac{4m}{1-\gm }sin^{1-\gm }(p)\; \; _{2}F_{1}
\left(1-\gm ,1;\frac{3-\gm }{2};\frac{1}{2}[1+cos(p)]\right). 
\eea
Now we introduce Kruskal-type coordinates, 
\bea 
U&=&2s\sqrt{A}\\
V&=&2p\sqrt{A}. 
\eea
The Light-Cone Gauge corresponds to the setting 
$\; \tau =U/4m\; $, that is to the choice \\ 
$\; \tau =-s\sqrt{A}/2m\; $. 
It is easy to verify, using Eqs (15), (18), (19), (21) and (24)-(25) 
that $\; ds^{2}=-dUdV\; $ in comformity 
with [8] and as it should be for Kruskal coordinates.

In Eq (22), the range of the parameters is $\; -\pi /2<s<\pi /2\; $ and also 
$\; 0<u<u_{0} \; $, where $\; u_{0}\equiv 4mC_{0}2^{1-\gm }/(1-\gm )\; $, 
(Appendix II).

From Eq (18), for this range of $\; s\; $, the function $u(s)$ as 
{\sl strictly increasing} and {\sl continuous} is {\sl invertible}. 
The variable u corresponds to the cosmic-time t and we are interested in the 
asymptotic regime $\; u\rightarrow u_{0}\; $where we shall seek steady state 
correlation functionals and stationary probability density functionals. 

We consider a relativistic string that falls freely, 
along the symmetry axis $\; \th =0\; $, 
towards the 
singular horizon. 
The wave equation in the free-fall frame is given by
\be
\left(\frac{\partial ^{2}}{\partial \tau ^{2}}
-\frac{\partial ^{2}}{\partial \sigma ^{2}}\right)X^{i}(\tau ,\sigma )=0. 
\ee
for the transverse coordinates $\; X^{i}(\tau ,\sigma )\; $.
In our case these are the spacelike transverse coordinates 
in the {\sl vicinity} 
of the hypersurface $\; \th =0\; $, because as it was proven 
in [12] with a proper choice of coordinates 
this is a two dimensional hypersurface for $\; \gm \geq 2\; $ 
while for our case of $\; 0<\gm <1\; $ it is a pure singularity. 
Therefore we consider a string which falls towards the singularity, 
with its transverse coordinates being normal to the symmetry axis 
$\th =0.$ \\
This can be written as 
\be
\left[\left(\frac{du}{d\tau }\right)^{2}
\frac{\partial ^{2}}{\partial u^{2}}+
\left(\frac{d^{2}u}{d\tau ^{2}}\right)\frac{\partial }{\partial u}
-\left(\frac{A}{4m^{2}}\right)
\frac{\partial ^{2}}{\partial \sigma ^{2}}\right]
X^{i}(u,\sigma )=0. 
\ee
We can express the derivatives of $\; u(s)\; $ with respect to 
$\; s,\; $ in terms of $u$ and this is \\
done in Appendix II. 

We proceed to the decomposition of the field and its
conjugate momentum, 
\bea
X^{i}(u,\sigma )&=&x^{i}(u,\sigma )+x^{i}_{f}(u,\sigma )\\
\dot{X}^{i}(u,\sigma )&\equiv &
\frac{\partial X^{i}(u,\sigma )}{\partial u}=v^{i}(u,\sigma )+
v^{i}_{f}(u,\sigma ),
\eea
into a slowly varying, classical part and a fast varying, quantum part. 
The quantum part is now expanded as a sum over modes in the free-fall 
frame, with a frequency cutoff that separates the fast modes from the slow 
ones, [8]  
\bea
x^{i}_{f}(u,\sigma )&=&\sum _{n=1}^{\infty }
W\left(n+\frac{\eps }{\tau }\right)\left[
\frac{c^{i}_{n}}{\sqrt{n}}x^{+}_{n}+
\frac{\tilde{c}^{i}_{n}}{\sqrt{n}}x^{-}_{n}+H.c.\right]\\
v^{i}_{f}(u,\sigma )&=&\sum _{n=1}^{\infty }
W\left(n+\frac{\eps }{\tau }\right)\left[
\frac{c^{i}_{n}}{\sqrt{n}}\dot{x}^{+}_{n}+
\frac{\tilde{c}^{i}_{n}}{\sqrt{n}}\dot{x}^{-}_{n}+H.c.\right]. 
\eea
Here $\; u=u(\tau ),\; $ $\; \eps >0\; $ is a constant, 
$\; W\; $is a Gaussian distribution function 
and the convention 
$\; x^{\pm }_{n}=\sqrt{\frac{\al ^{'}}{2}}e^{-in(\tau \pm \sigma )}\; $
is used. The commutation relations for the field operators are 
$\; [c^{i}_{n},(c^{j}_{m})^{\dagger }]=\delta _{mn}\delta ^{ij}\; $ 
and similar for the tilded operators.

From these relations we obtain
\bea
\dot{x}^{i}_{f}(u,\sigma )&=&\sum _{n=1}^{\infty }
W\left(n+\frac{\eps }{\tau }\right)\left[
\frac{c^{i}_{n}}{\sqrt{n}}\dot{x}^{+}_{n}+
\frac{\tilde{c}^{i}_{n}}{\sqrt{n}}\dot{x}^{-}_{n}+H.c.\right]
-\eta ^{i}(u,\sigma )\\
\eta ^{i}(u,\sigma )&=&
\left(\frac{du}{d\tau }\right)^{-1}
\left(\frac{\eps }{\tau ^{2}}\right)
\sum _{n=1}^{\infty }
W^{'}\left(n+\frac{\eps }{\tau }\right)\left[
\frac{c^{i}_{n}}{\sqrt{n}}x^{+}_{n}+
\frac{\tilde{c}^{i}_{n}}{\sqrt{n}}x^{-}_{n}+H.c.\right]. 
\eea
Also in the same manner we have
\bea
\dot{v}^{i}_{f}(u,\sigma )&=&\sum _{n=1}^{\infty }
W\left(n+\frac{\eps }{\tau }\right)\left[
\frac{c^{i}_{n}}{\sqrt{n}}\ddot{x}^{+}_{n}+
\frac{\tilde{c}^{i}_{n}}{\sqrt{n}}\ddot{x}^{-}_{n}+H.c.\right]
-\xi ^{i}(u,\sigma )\\
\xi ^{i}(u,\sigma )&=&
\left(\frac{du}{d\tau }\right)^{-1}
\left(\frac{\eps }{\tau ^{2}}\right)
\sum _{n=1}^{\infty }
W^{'}\left(n+\frac{\eps }{\tau }\right)\left[
\frac{c^{i}_{n}}{\sqrt{n}}\dot{x}^{+}_{n}+
\frac{\tilde{c}^{i}_{n}}{\sqrt{n}}\dot{x}^{-}_{n}+H.c.\right].  
\eea
Substitution into the field equations for the $\; X^{i}\; $
results in to two coupled equations for the long-wavelength
fields, 
\bea
\dot{x}^{i}&=&v^{i}+\eta ^{i}\\
\dot{v}^{i}&=&-\frac{u^{''}}{(u^{'})^{2}}v^{i}
+\frac{1}{(u^{'})^{2}}\left(\frac{A}{4m}\right)
\frac{\partial ^{2}}{\partial \sigma ^{2}}x^{i}
+\xi ^{i}. 
\eea
Here $\; u^{''}, u^{'}\; $are the derivatives of $u$ with respect to 
$\; (s)\; $, expressed in terms of $u$. 
Substituting into the second equation we obtain
\be 
\dot{v}^{i}(u,\sigma )=-H_{1}(u)
v^{i}(u,\sigma )+H_{2}(u)
\frac{\partial ^{2}}{\partial \sigma ^{2}}x^{i}(u,\sigma )+\xi ^{i}. 
\ee 
The form of the functions $\; H_{1}(u), H_{2}(u)\; $is given 
by direct comparison with Eq (37). 
The spatial derivative term becomes negligible with respect to 
the first term on the r.h.s. of Eq (38), as $\; u\rightarrow u_{0-}\; $, 
that is, when the string approaches the singularity at $\; r=2m\; $. 
An additional supporting argument for this is given in 
Appendix III. 
Procceding as in [8] we finally obtain 
\bea
<\eta ^{i}(1)\eta ^{j}(2)>&=&
\frac{\al ^{'}}{2}
\delta ^{ij}
\left(\tau \frac{du}{d\tau }\right)^{-1}
\delta [u(\tau _{1})-u(\tau _{2})]\cdot \nn \\
& &\cdot exp\left[-\frac{\beta ^{2}}{4}\left(
\frac{\eps \Delta \sigma }{\tau }\right)^{2}\right]
cos\left(\frac{\eps \Delta \sigma }{\tau }\right). 
\eea
This correlation function shows that the string bits undergo 
Brownian motion since the correlation of two string bits is 
practically zero outside the {\sl correlation length} 
$\; \Delta \sigma =|\tau |/\bt \eps \; $. 

The rest of the correlation functions can be computed in the same way. 
We have
\bea
\dot{x}_{n}^{\pm }\equiv \frac{\partial }{\partial u}x_{n}^{\pm }=
-in\left(\frac{du(\tau )}{d\tau }\right)^{-1}x_{n}^{\pm }=
-i\left(\frac{\eps }{\tau }\right)
\left(\frac{du(\tau )}{d\tau }\right)^{-1}x_{n}^{\pm }, 
\eea
because the main contribution comes from $\; n\simeq (\eps /\tau )\; $.
So
\bea
<\xi ^{i}(1)\xi ^{j}(2)>=
-\left[\frac{\eps }{\tau }
\left(\frac{du(\tau )}{d\tau }\right)^{-1}\right]^{2}
<\eta ^{i}(1)\eta ^{j}(2)> \\
<\xi ^{i}(1)\eta ^{j}(2)>=
-\left[\frac{i\eps }{\tau }
\left(\frac{du(\tau )}{d\tau }\right)^{-1}\right]
<\eta ^{i}(1)\eta ^{j}(2)>. 
\eea
Now these correlator functions are negligible in 
comparison to Eq (39) as the string approaches the singularity, 
that is as $\; u\rightarrow u_{0-}\; $. 
This is evident from Eq (83). \\
\section*{IV. The Diffusion Process and the Fokker-Planck equation}\ 
For the case of Schwarzschild spacetime $\; (\gm =1)\; $we obtain the 
same results as in [8], because the term 
$\; (\tau \frac{du}{d\tau })^{-1}\; $ in Eq (39) reduces to $\; (1/4M)\; $. 
For the present case the coefficient of the correlator is the diffusion 
coefficient for the process. It is given by 
\be 
\zeta ^{2}_{1}\equiv \left(\frac{\al ^{'}}{8m}\right)
\left(\frac{2^{\gm +1}}{\pi }\right)
\left(\frac{u}{u_{0}}\right)^{\gm /1-\gm }
\left[1-\left(\frac{u}{u_{0}}\right)^{2/1-\gm }\right]^{\gm /2}. 
\ee 
By the {\sl fluctuation-dissipation} theorem the temperature of the 
process assigned by an asymptotic observer is inversely proportional 
to the diffusion coefficient [18]. As $\; u\rightarrow u_{0-}\; $ 
the temperature grows without limit.

From Eqs (41)-(42) one concludes that the diffusion process in momentum 
space is negligible and can be omitted with respect to the diffusion 
process in ordinary space. We will however retain this also and consider 
that it evolves with a diffusion coefficient $\; \zeta _{2}\ll \zeta _{1}\; $.
We rewrite the two Langevin equations  as 
\bea 
\frac{\partial x^{i}(u,\sigma )}{\partial u}&=&v^{i}+\eta ^{i}\\ 
\frac{\partial v^{i}(u,\sigma )}{\partial u}&=&-H_{1}(u)v^{i}+\xi ^{i}. 
\eea 
The Fokker-Planck equation for the probability density 
$\; \Phi =\Phi (x^{i},v^{i};u)\; $, corresponding to the above set of 
equations is given by [19]
\bea 
\frac{\partial \Phi }{\partial u}&=&F\Phi \nn \\
F\equiv \sum _{i}\left[\zeta _{1}^{2}
\frac{\partial ^{2}}{\partial (x^{i})^{2}}\right.&+&
\left.\zeta _{2}^{2}\frac{\partial ^{2}}{\partial (v^{i})^{2}}-
v^{i}\frac{\partial }{\partial x^{i}}+
H_{1}(u)\frac{\partial }{\partial v^{i}}(v^{i} )\right]. 
\eea 
The probability density can be normalized for all "times" if it is 
normalized once. This is because we have 
\bea 
\frac{d}{du}\int dx^{i}dv^{i}\Phi (x^{i},v^{i};u)=
\int dx^{i}dv^{i}\frac{\partial }{\partial u}\Phi (x^{i},v^{i};u)=\nn \\
=\int dx^{i}dv^{i}F\Phi (x^{i},v^{i};u)=0, 
\eea 
which vanishes under proper boundary conditions, 
because the action of the operator $\; F\; $can be written 
as a divergence with respect to the space variables. We denote 
collectively the space variables $\; (x^{i})\; $ by (q) and 
$\; (v^{i})\; $ by (v). \\
Writing $\; F=F_{1}+H_{1}(u)F_{2}\; $ 
where it is evident the content of the two terms, we can write formally 
the solution of the Fokker-Planck equation as 
\be 
\Phi (q,v;u)=exp[uF_{1}+G(u)F_{2}]\delta (q)\delta (v), 
\ee 
with 
\be 
G(u)\equiv 
\int_{0}^{u}du^{'}H_{1}(u^{'}). 
\ee 
Here, the usual convention, $\; 
\delta (q)=\frac{1}{2\pi }\int_{-\infty }^{+\infty }d\rho e^{iq\rho },\; \; 
\delta (v)=\frac{1}{2\pi }
\int_{-\infty }^{+\infty }d\bar{\rho }e^{iv\bar{\rho }}\; $
is used. 
We were able to solve completely the time-dependent Fokker-Planck 
Eq (46) without any sort of 
approximation. The solution is given by 
\bea 
\Phi (q,v;u)&=&exp\left[-\frac{[q-Q_{2}(u)v]^{2}}{Q_{1}(u)}\right]U(v;u)\\ 
U(v;u)&=&\Psi (u)exp\left[-\frac{v^{2}}{4\zeta _{2}^{2}Q_{2}}
(1-Q_{2}^{'}+H_{1}(u)Q_{2})\right], 
\eea 
where the the three functions $\; Q_{1}(u), Q_{2}(u), \Psi (u)\; $ 
satisfy 
\bea 
Q_{1}^{'}(u)&=&4\zeta _{1}^{2}+4\zeta _{2}^{2}Q_{2}^{2}\\
(f+H_{1})^{'}&+&(f^{2}-H_{1}^{2})+
2f\left[\frac{Q_{1}^{'}-4\zeta _{1}^{2}}{Q_{1}}\right]=0\\
f&\equiv &\left(\frac{1-Q_{2}^{'}}{Q_{2}}\right)\\
\Psi ^{'}(u)&=&-\frac{2\zeta _{1}^{2}}{Q_{1}}-
\frac{1}{2}(f+H_{1})-
\frac{2\zeta _{2}^{2}Q_{2}^{2}}{Q_{1}}+H_{1}.
\eea 
Prime denotes ordinary differentiation with respect to the (u).
This set of equations cannot be solved in closed form for a generic 
choice of the function $\; H_{1}(u)\; $. In Appendix III, the 
asymptotic form of $\; H_{1}(u)\; $ is given as 
$\; u\rightarrow u_{0-}.\; $ For this choice the solution is 
given by 
\bea 
f&=&-H_{1}\\ 
Q_{1}&=&4\zeta _{1}^{2}u\\ 
Q_{2}&=&1-exp\left[-\kp _{0}(\frac{1+\gm }{2})^{-1}
(u_{0}-u)^{\frac{1+\gm }{2}}
\right], 
\eea 
with the only approximation made up to here, 
is that as $\; u\rightarrow u_{0-},\; $ 
$\; Q_{2}(u)\simeq 0\; $so that the second term on the r.h.s. 
of Eq (52) can be neglected, as being of the second order,  
in the first order approximation case. 
Neglecting, for the same argument, 
the third term in the r.h.s. of Eq (55), one can integrate 
it, obtaining 
\be 
\Psi (u)=ln\frac{1}{\sqrt{u}}-\frac{2\kp _{0}}{(1+\gm )}
(u_{0}-u)^{(1+\gm )/2}. 
\ee 
From Eq (50), we now see, that we 
can have a product of two thermal distributions, 
provided that the cross term in the exponential is constant, 
or can be neglected. 
Denoting this term as  
\be 
T(u)\equiv \frac{2\kp _{0}}{(1+\gm )}
\frac{(u_{0}-u)^{(1+\gm )/2}}{2\zeta _{1}^{2}u}qv
\ee 
and following the same procedure as in Appendix III, we find that 
the asymptotic form of Eq (43), as $\; u\rightarrow u_{0-}\; $ 
is given by $\; \zeta _{1}^{2}\propto (u_{0}-u)^{\gm /2}\; $. 
Therefore in Eq (60) the overall dependence is given by 
$\; T(u)\propto (u_{0}-u)^{1/2}\; $. 
Now we assume that the {\sl Uncertainity Principle} relation 
holds for the phase space variables $\; qv=\hbar ,\; \; \hbar =$constant. 
Then from Eq (60) the cross term vanishes and we have a product of two 
thermal distribution functions. 

\section*{V. Numerical Estimates}\
There exist a variety of physically interesting systems that 
one could refer to, in order to ascertain that the range of 
parameters under study, is acceptable. Using 
{\sl geometical units} we set $\; c=G=K=1\; $ with  
$\; c=2.99\times 10^{10}\; $cm/sec=1 and 
$\; G/c^{2}=0.74\times 10^{-28}\; $cm/gr=1 for the conversion factors [20]. 

For cosmic strings, one can consider two extreme cases. A string that 
originates in symmetry breaking at a mass scale 
$\; \Lambda \simeq 10^{16}\; $GeV (the scale of GUT's) has a mass per 
unit length of the order $\; 10^{22}\; $gr/cm [21]. 
On the other side strings arising in the electroweak scale 
$\; \Lambda \simeq 1 \; $TeV has a mass per unit length of 
the order $\; 10^{-6}\; $gr/cm. Such a string with the length of 
a galaxy (radius$\; 10^{22}$cm) would have a mass of the order of 
$\; 10^{16}\; $gr and it would be unobservable by gravitational 
interactions. For the first case we obtain 
$\; \gm =0.74\times 10^{-6}\ll 1\; $ while for the second 
$\; \gm =0.74\times 10^{-34}\ll 1\; $. 

A more realistic prospect comes from the fact that the mass of the Sun 
is \\
$\; M_{\odot }=1.989\times 10^{33}$gr$=1.477\times 10^{5}$cm. 
In order that we get a value of 
\be 
\gm =\frac{1M_{\odot }}{2\al }=\frac{1}{2}, 
\ee 
we must have $\; \al =1.447\times 10^{5}$cm, which compared with 
the Sun radius $\; R_{\odot }\simeq 7\times 10^{10}$cm gives a ratio 
of the order of $\; 10^{-5}\; $. 

On the Galactic scale, the most extreme case corresponds to 
the masses of central regions [22-23]. Here we have 
$\; M_{BH}\simeq 10^{9}M_{\odot }=1.477\times 10^{14}$cm, whereas 
the central region extends over the scale of 
$\; R\simeq 0.1$pc$=3.086\times 10^{17}$cm. This yields 
$\; \gm \simeq 0.3\times 10^{-3}.\; $ Therefore in all cases, the range 
of parameters under study is the relevant one, for almost all 
astrophysically interesting systems. 

\section*{VI. Discussion}\
The concept of the {\sl Stretched Horizon} has been introduced 
recently, that thermalizes the quantum states of the infalling matter 
towards the black hole and eventually reemits them in the form of the 
thermal Hawking radiaton. This prevents the loss of information 
inside the black hole Horizon during the gravitational collapse so that 
there exists at this first level of analysis no conflict between the 
notion of gravitational collapse and the unitary evolution of states 
in quantum theory. This is achieved by considering that matter is in the 
form of strings, that undergo a diffusion process in the space of the 
transverse variables.\\ 
In our case the spacetime under consideration 
contains a singular event horizon for $\; \th =0\; $ 
and $\; \gm \geq 2\; $, 
while for other values of $0<\gm <1$, 
a naked singularity along all the directions. Therefore 
there doesn't exist a Horizon and no natural distinction between an 
asymptotic and a 
free-falling observer. Even more we have chosen to describe the 
process of {\sl Branching Diffusion} in terms of the coordinate-based 
observer  that in principle, can approach arbitrary 
close to the singularity. If however we invoke the fact that 
strings are quantum 
objects, this provides the mechanism for obtaining thermal spectrum for the 
diffusion processes of string bits in both the configuration and the 
momentum space of the transverse variables of the string. Therefore we 
can associate with it a temperature. This has been achieved for the 
particular class of static models that have $\; 0<\gm <1\; $. Although we 
considered the case of the symmetry axis where there exists the singularity, 
by conntinuity our arguments are valid in the vicinity of $\; \th =0\; $. 
Therefore in the case 
of the naked singularity we also have thermalization process. 

Now, although it seems that this temperature increases without limit, 
there exists a very interesting fact here. It has to do with the fact that 
practically the string is somehow thermalized at some finite 
distance away from the singularity, 
in whatever quantum state it had been initially. The string bits become 
practically uncorrelated over a spatial distance of the same order of 
magnitude as the quantum correlation length. 
In principle then, very little or no information, can get lost on 
the singular event horizon, or can escape from it. 

One can reverse this argument and examine 
whether this is some form of {\sl Cosmic Censorship}.
This is enhanced by the fact that this is a static spacetime 
and there exists the invariance under coordinate-time reversal 
$\; t\leftrightarrow (-t)\; $. 
Also in the same spirit, it would be interesting, if one could 
address the same problem for 
membranes or other extended objects of modern string theory. 
Work along these lines is is 
progress and will be reported promptly. 

\section*{Acknowledgements.} 
The authors would like to thank Professor Louis Witten for 
valuable discussions. 
One of us (A.K.) would like to thank the Greek
State Scholarships Foundation (I.K.Y.), for the financial support 
during this work. This work is partially supported by the Scientific 
Programme PENED 1768 (GREECE).

\section*{Appendix I}
We give below, for the sake of completeness, the 
non-zero components of the Riemann curvature tensor 
for the $\; (\gamma )\; $metric. 
They are consistent with the conditions 
\bea
R_{\alpha \beta \gamma \delta }=R_{[\alpha \beta ][\gamma \delta ]}&=&
R_{\gamma \delta \alpha \beta }\nn \\
R_{[\alpha \beta \gamma \delta ]}&=&0. 
\eea
The expressions are considerably complicated. With the help of 
the definitions
\bea 
{\cal A}&=&\left(1-\frac{2m}{r}\right)\\
{\cal B}&=&\left(1-\frac{2m}{r}+\frac{m^{2}}{r^{2}}sin^{2}\theta \right),\\
\eea
we have 
\be
R_{\phi t\phi t}=\frac{1}{2}\gm \left(\frac{{\cal A}_{,r}}{{\cal A}}
\right)\left(\frac{{\cal B}}{{\cal A}}\right)^{\gm ^{2}-1}
{\cal A}^{\gm }
\left[{\cal A}-\frac{1}{2}(\gm -1)r{\cal A}_{,r}\right]
rsin^{2}\theta 
\ee
\be
R_{\th t\th t}=\frac{1}{2}\gm \left(\frac{{\cal A}_{,r}}{{\cal A}}
\right)
{\cal A}^{\gm }
\left[\frac{1}{2}\gm (\gm -1)r^{2}{\cal A}_{,r}
-\frac{1}{2}\gm (\gm ^{2}-1)r^{2}{\cal A}
\left(\frac{{\cal B}_{,r}}{{\cal B}}\right)-
r{\cal A}\right]
\ee
\be
R_{rt\th t}=\frac{1}{4}\gm (\gm ^{2}-1)\left(
\frac{\cA _{,r}}{\cA }\right)
\left(\frac{\cB _{,\th }}{\cB }\right)
\cA ^{\gm }
\ee 
\bea
R_{rtrt}=\frac{1}{2}\cA ^{\gm }
\left[ \gm \left(\frac{\cA _{,rr}}{\cA }\right)-
\frac{1}{4}\gm (\gm ^{2}-1)
\left(\frac{\cA _{,r}}{\cA }\right)
\left(\frac{\cB _{,r}}{\cB }\right)-
\frac{1}{4}\gm (\gm -1)^{2}
\left(\frac{\cA _{,r}}{\cA }\right)^{2}\right]
\eea
\bea
R_{\th \phi \th \phi }=\left. 
\frac{r^{2}}{\cA ^{\gm -1}}\right[
&-&\frac{1}{2}(\gm ^{2}-1)
\left(\frac{\cB _{,\th }}{\cB }\right)sin\th cos\th +
(1+\cA )sin^{2}\th +\nn \\
&+&\frac{1}{4}\gm (\gm -1)^{2}
\left(\frac{\cA _{,r}}{\cA }\right)r^{2}\cA _{,r}sin^{2}\th -\nn \\
&-&\frac{1}{4}(\gm -1)(\gm ^{2}-1)
\left(\frac{\cB _{,r}}{\cB }\right)r^{2}\cA _{,r}sin^{2}\th -
\frac{1}{2}(\gm -1)r\cA _{,r}sin^{2}\th -\nn \\
&-&\frac{1}{2}\gm (\gm -1)r\cA _{,r}sin^{2}\th +
\left.\frac{1}{2}(\gm ^{2}-1)
\left(\frac{\cB _{,r}}{\cB }\right)r\cA sin^{2}\th \right]
\eea
\bea
R_{r\phi \th \phi }=\left.
\frac{1}{\cA ^{\gm }}\right[
\frac{1}{2}(\gm ^{2}-1)r^{2}\cA _{,r}sin\th cos\th &-&
\frac{1}{2}(\gm ^{2}-1)
\left(\frac{\cB _{,r}}{\cB }\right)r^{2}\cA sin\th cos\th -\nn \\
-2r\cA sin\th cos\th &+&\frac{1}{4}(\gm -1)(\gm ^{2}-1)
\left(\frac{\cB _{,\th }}{\cB }\right)r^{2}\cA _{,r}sin^{2}\th -\nn \\
&-&\frac{1}{2}(\gm ^{2}-1)
\left.\left(\frac{\cB _{,\th }}{\cB }\right)r\cA sin^{2}\th \right]
\eea
\bea 
R_{r\th r\th }=\frac{1}{2}\cA ^{-\gm }
\left(\frac{\cA }{\cB}\right)^{\gm ^{2}-1}
\left[\gm (1-\gm )r^{2}\cA _{,rr}+(\gm ^{2}-1)
\left(\frac{\cA }{\cB }\right)r^{2}\cB _{,rr}+\right. \nn \\
+(\gm ^{2}-1)\left(\frac{\cB _{,\th \th }}{\cB }\right)-
r\cA _{,r}(\gm ^{2}-\gm +1)+\nn \\
+\frac{1}{2}(\gm ^{2}-1)\left(\frac{\cB _{,r}}{\cB }\right)r^{2}\cA _{,r}
+\frac{1}{2}\gm (\gm -1)\left(\frac{\cA _{,r}}{\cA }\right)r^{2}\cA _{,r}-\nn \\
\left.-(\gm ^{2}-1)\left(\frac{\cB _{,r}}{\cB }\right)^{2}r^{2}\cA +
(\gm ^{2}-1)\left(\frac{\cA }{\cB }\right)r\cB _{,r} -
(\gm ^{2}-1)\left(\frac{\cB _{,\th }}{\cB }\right)^{2}\right]
\eea
\bea 
R_{r\phi r\phi }=\frac{1}{2}\cA ^{-\gm }
\left[(\gm -1)r^{2}\cA _{,rr}\right.
&+&(\gm ^{2}-1)\left(\frac{\cB _{,\th }}{\cB }\right)cot\th -
2r\cA _{,r}-\nn \\
-\frac{1}{2}\gm ^{2}(\gm -1)\left(\frac{\cA _{,r}}{\cA }\right)
r^{2}\cA _{,r}
&-&(\gm ^{2}-1)\left(\frac{\cB _{,r}}{\cB }\right)r\cA +\nn \\
+(\gm ^{2}+\gm -1)\left(\frac{\cA _{,r}}{\cA }\right)r\cA 
&+&\frac{1}{2}(\gm -1)(\gm ^{2}-1)\left(\frac{\cB _{,r}}{\cB }\right)
r^{2}\cA _{,r}\left.\right]sin^{2}\th .
\eea

\section*{Appendix II}
We consider Eq (18) and the case $\; 0<\gm <1\; $. 
Setting $\; u(s)=-4mcos^{1-\gm }(s)f(s)\; $ one finds that f(s) satisfies
\be
cos(s)f^{'}(s)+(\gm -1)sin(s)f(s)+1=0. 
\ee
Setting $\; h=sin(s)\; $ and 
subsequently $\; h=(2t-1)\; \; \; (0<t<1)\; $one finally
gets
\be
2t(1-t)\frac{df}{dt}+(1-\gm )(1-2t)f(t)+1=0. 
\ee
It can be shown that the following is a solution to this equation [17],
\be 
f(t)=-\frac{1}{1-\gm }\; _{2} F_{1}(1-\gm ,1;\frac{3-\gm }{2};t), 
\ee 
by expanding in a power series f(t) and substituting into Eq (75).
Using \\
MATHEMATICA, Eq (18) is integrated as 
\bea 
u(s)&=&\frac{4m}{1-\gm }\frac{cos^{1-\gm }(s)}{\sqrt{sin^{2}(s)}}
\; \; _{2}F_{1}
\left(\frac{1-\gm }{2},\frac{1}{2};\frac{3-\gm }{2};cos^{2}(s)\right)sin(s)
\eea 
These are exactly equivalent due to a property of hypergeometric 
functions (see [17], p.561, 15.3.30). 
In the same way one integrates Eq (19).\\ 
We use now two well known properties of the hypergeometric 
function to obtain the limit $\; t\rightarrow 1_{-}\; $. 
\bea 
\; _{2}F_{1}(a,b;c;t)&=&(1-t)^{c-a-b}\; _{2}F_{1}(c-a,c-b;c;t)\nn \\
\; _{2}F_{1}(a,b;c;1)&=&\frac{\Gm (c)\Gm (c-a-b)}{\Gm (c-a)\Gm (c-b)}
\; \; \; \; (c-a-b>0,\; \; b>0). 
\eea 
So we obtain 
\bea 
\; _{2} F_{1}(1-\gm ,1;\frac{3-\gm }{2};t)&=&
(1-t)^{-(1-\gm )/2}
\; _{2} F_{1}(-\frac{1+\gm }{2},\frac{1-\gm }{2};\frac{3-\gm }{2};t\simeq 1)=
\nn \\
&=&(1-t)^{-(1-\gm )/2}
\frac{\Gm (\frac{3-\gm }{2})\Gm (\frac{1-\gm }{2})}{\Gm (1-\gm )}\nn \\
&\equiv &C_{0}(1-t)^{-(1-\gm )/2}. 
\eea 
For t=1, which corresponds to $\; s=\pi /2\; $, we obtain that 
\be 
u_{0}\equiv u(s=\pi /2)=\frac{4mC_{0}}{1-\gm }2^{1-\gm }. 
\ee 
Therefore in the vicinity of $\; t\simeq 1\; $, 
we have 
\be 
u(s)=\frac{4m}{1-\gm }C_{0}(4t)^{(1-\gm )/2}, 
\ee 
so we can invert 
Eq (22) to obtain 
\be 
sin(s)=\left[2\left(\frac{u}{u_{0}}\right)^{2/(1-\gm )}-1\right] 
\ee 
and 
\bea 
\frac{du}{ds}&=&\frac{4m}{2^{\gm }}\left(\frac{u}{u_{0}}\right)^{-\gm /(1-\gm )}
\left[1-\left(\frac{u}{u_{0}}\right)^{2/(1-\gm )}\right]^{-\gm /2} \nn \\ 
\frac{d^{2}u}{ds^{2}}&=&
\frac{4m\gm }{2^{\gm +1}}\left(\frac{u}{u_{0}}\right)^{-(\gm +1)/(1-\gm )}
\left[1-\left(\frac{u}{u_{0}}\right)^{2/(1-\gm )}\right]^{-(\gm +1)/2} 
\left[2\left(\frac{u}{u_{0}}\right)^{2/(1-\gm )}-1\right]. 
\eea

\section*{Appendix III}
From Appendix II we obtain 
\be 
\lim_{u\rightarrow u_{0-}}\frac{H_{2}(u)}{H_{1}(u)}=0 
\ee 
and the spatial derivative term is negligible. 
Alternatively, the vector 
$\; \Xi ^{i}\equiv \partial x^{i}/\partial \sigma \; $\\ 
is the separation vector of two points on the string, obeying the 
{\sl geodesic deviation} equation 
\be 
\frac{D^{2}\Xi ^{i}}{d\tau ^{2}}+R^{i}_{jkl}u^{j}\Xi ^{k}u^{l}=0. 
\ee 
For $\; i=\phi \; $, we have from Appendix I, that the second term 
on the l.h.s. of Eq (66) vanishes. Then it is solved as 
$\; \Xi ^{\phi }(\tau ,\sigma )=\Xi _{0}(\sigma )\tau +\Xi _{1}(\sigma )\; $.
So the spatial derivative term is proportional to the proper time 
$\; (\tau )\; $ which remains finite on the singular horizon. \\
We proceed now to obtain the asymptotic form of 
$\; H_{1}(u)\; $, as $\; u\rightarrow u_{0-}\; $. 
Combining Eqs (37), (38) and (83) we have 
\be 
H_{1}(u)=\left(\frac{\gm }{2m}\right)
\left(\frac{1}{2^{\gm }}\right)
\left[1-\left(\frac{u}{u_{0}}\right)^{2/(1-\gm )}\right]^{-(1-\gm )/2}.
\ee 
Denoting, 
\be 
0<\eps \equiv \left(\frac{u_{0}-u}{u_{0}}\right)\ll 1,
\ee 
we have that 
\bea 
\left[1-\left(\frac{u}{u_{0}}\right)^{2/(1-\gm )}\right]^{-(1-\gm )/2}=
\left[1-(1-\eps )^{2/(1-\gm )}\right]^{-(1-\gm )/2}\simeq 
\left(\frac{1-\gm }{2\eps }\right)^{(1-\gm )/2},
\eea 
so that 
\bea 
H_{1}(u)&=&\frac{\kp _{0}}{(u_{0}-u)^{(1-\gm )/2}}\nn \\
\kp _{0}&=&\left(\frac{\gm }{2m}\right)
\left(\frac{1}{2^{\gm }}\right)
\left(\frac{(1-\gm )u_{0}}{2}\right)^{(1-\gm )/2}.
\eea


\begin{thebibliography}{99}

\bibitem{}
M.Green, J.Schwarz, E.Witten, {\sl Superstring Theory.}
   Cambridge University Press,
Cambridge 1987.
\bibitem{}
H. J. de Vega and N. S\'{a}nchez {\sl String Theory in 
Cosmological Spacetimes},
Lectures 
at the Erice School, September (1994).
\bibitem{}
H. J. de Vega {\sl Strings in Curved spacetimes}, 
Lectures at the Erice 
School, June (1992).
\bibitem{}
G. 't Hooft, {\sl The Black Hole Interpretation of String Theory},
Nucl. Phys. {\bf B335} 
(1990) 138-154.
\bibitem{} 
L. Susskind, L. Thorlacius and J. Uglum, {\sl The stretched horizon 
and black hole complementarity}, Phys. Rev. D 48, 3743-3760 (1993).
\bibitem{}
L. Susskind, {\sl String Theory and the Principle of Black Hole 
Complementarity}, Phys. Rev. Lett. 71, 2367-2368 (1993).
\bibitem{}
L. Susskind and J. Uglum, {\sl Black hole entropy in canonical
quantum gravity and 
superstring theory}, Phys. Rev. D 50, 2700-2711 (1994).
\bibitem{}
A. Mezhlumian, A. Peet and L. Thorlacius, 
{\sl String thermalization at a black hole 
horizon}, Phys. Rev. D 50, 2725-2730 (1994).
\bibitem{}
Kenji Hotta, {\sl The Information Loss Problem of Black Hole and the 
First Order
Phase Transition in String theory}, preprint, hep-th/9705100 (1998) 
\bibitem{}
C. W. Misner, K. S. Thorne, J. A. Wheeler, {\sl Gravitation},
Freeman, San Francisco, (1973).
\bibitem{}
C. Duncan, F. Paul Esposito and Scott Lee, Phys. Rev. D {\bf 17},
404 (1978).
\bibitem{}
D. Papadopoulos, B. Stewart and L. Witten, 
{\sl Some properties of a particular static, axially symmetric 
space-time}, Phys. Rev. D  {\bf 24}, 320-326 (1981).
\bibitem{}
D. Papadopoulos, {\sl Effects of a Coordinate Mapping on a Particular 
Class of Static, Axially Symmetric Solutions to the Einstein Equations},
Lettere Al Nuovo Cimento, vol {\bf 44} N. 7 497-502.
\bibitem{} 
Jiri Bicak, D. Lynden-Bell and Joseph Katz, 
{\sl Relativistic disks as sources of 
static vacuum spacetimes}, Phys. Rev. D {\bf 47}, 4334-4343 (1993).
\bibitem{} 
J. P. S. Lemos, {\sl Remarkable properties of the limiting
counter-rotating disc}, Mon. Not. R. astr. Soc. (1988) {\bf 230}, 451-456. 
\bibitem{}
G. T. Horowitz and S. F. Ross, {\sl Properties of Naked Black Holes}, 
preprint, hep-th/9709050 (1997)
\bibitem{}
M. Abramowitz and I. A. Stegun, {\sl Handbook of Mathematical Functions}, 
Dover Publications, Inc., N.Y., (1970) 
\bibitem{}
D. R. Nicholson, {\sl Introduction to Plasma Theory}, 
J. Wiley and Sons, Inc., N.Y., (1983) 
\bibitem{}
M. Namiki, {\sl Stochastic Quantization}, Lecture Notes in Physics m9, 
Springer Verlag, N.Y., (1992)
\bibitem{}
C.W. Misner, K.S. Thorne, J.A. Wheeler, {\sl Gravitation}, 
W.H. Freeman, 1973. 
\bibitem{}
E. Witten, {\sl Superconducting Strings}, Nucl. Phys. B249 (1985) 557-592. 
\bibitem{}
M. Miyoshi, et al., Nature, Vol 373, (1995) 127-129.
\bibitem{}
L.J. Greenhill, et al., Astrophys. J. 440, 619-627, (1995). 

\end{thebibliography}
\end{document}